\documentclass[epj]{svjour}
\usepackage{latexsym}
\usepackage{graphics}
\usepackage{amssymb}
\usepackage{xcolor}
\usepackage{cite}
\usepackage{soul}

\begin{document}
\title{Topology dependent payoffs can lead to escape from prisoner's dilemma}
\author{Saptarshi Sinha\inst{1} \and Deep Nath\inst{1} \and Soumen Roy\inst{1} % etc
\thanks{{e-mail:} soumen@jcbose.ac.in}%
}                     % Do not remove
\institute{Department of Physics, Bose Institute, 93/1 Acharya Prafulla Chandra Road, Kolkata 700009, India}
\date{Received: \today}
\abstract{
The maintenance of cooperation in the presence of spatial restrictions has been studied extensively.  It is well-established that the underlying graph topology can significantly influence the outcome of games on graphs. Maintenance of cooperation could be difficult, especially in the absence of spatial restrictions. The evolution of cooperation would naturally depend on payoffs. However, payoffs are generally considered to be invariant in a given game. A natural yet unexplored question is whether the topology of the underlying structures on which the games are played, possesses no role whatsoever in the determination of payoffs. Herein, we introduce the notion of cooperator graphs and defector graphs as well as a new form of game payoff, which is weakly dependent on the underlying network topology. These concepts are inspired by the well-known microbial phenomenon of quorum sensing. We demonstrate that even with such a weak dependence, the fundamental game dynamics and indeed the very nature of the game may be altered. Such changes in the nature of a game have been well-reported in theoretical and experimental studies.
 % end of PACS codes
} %end of abstract
\maketitle
\section{Introduction}
We can readily appreciate the altruistic nature of the living world in our surroundings \cite{cheney2011extent,sachs2012origins}. However, the charitable nature of cooperators  can be exploited easily by defectors or free riders. A few mechanisms which support the evolution of cooperators with free riders of higher fitness are known \cite{PercBiosys2010}. Evolutionary game theory explains  evolutionary outcomes as a deterministic and frequency-dependent steady state of the population using game rules   \cite{Szabo2007,Comolli2014}. Before its advent, evolution was considered as a time-dependent stochastic process \cite{smith}. Players are constrained to adhere strictly to pure strategies in evolutionary games, due to their genetic makeup. This can either be cooperation, $C$, or defection, $D$. Each interaction results in a definite payoff for both players. When two cooperators  interact with each other, they receive a reward ${\cal R}$. Similarly, interactions between two defectors results in punishment, ${\cal P}$. On the other hand, the cooperator receives a sucker's payoff, ${\cal S}$, and the defector receives a temptation, ${\cal T}$, when a cooperator and a defector  interact. 

An individual accumulates its payoff from various interactions with its neighbors. Two interpretations exist in literature on the role of payoff in fitness determination: (1) the selectionist approach \cite{santos2005scale}, and, (2) the naturalist approach \cite{Ohtsuki2006, Szolnoki2009}.  The former considers game payoff as the sole deciding factor of a player's fitness. However, according to the latter, game payoff is only one of the several deciding factors. 
The prisoner's dilemma (PD) is  widely studied in evolutionary game theory, with payoff values satisfying ${\cal T} > {\cal R} > {\cal P} > {\cal S}$  \cite{Szabo2007}. In PD, defection is always the best strategy regardless of the initial configuration of the population. Since cooperation is not favored in a typical PD scenario, a proper understanding of the maintenance of cooperation in such scenarios has been the prime focus for many evolutionary biologists and ecologists \cite{Gore2009}. PD game on different types of graph structures such as Barabasi-Albert (BA) network, small-world network, random-regular network etc has been well investigated \cite{Szabo2007, santos2005scale, Sinha2019}.  The outcome of the game is sensitive to the underlying graph structure. 

Till date, most of the research on evolutionary game theory has been primarily concentrated on structured and spatially restricted populations \cite{nowak1992evolutionary, Perc2008, Roca2009, perc2017statistical}. Stimulating correspondences between spatial evolutionary game theory and nonequilibrium phase transitions have provided a fresh and engaging view to the concept of universality classes. Indeed, evolutionary game theory in structured populations exhibits critical phase transitions that lie in the universality class of directed percolation on square lattices. Critical phase transitions similar to mean-field-type transitions on random regular graphs and regular small world networks have also been observed in evolutionary games on structured populations. Such spatial restrictions constrain players to interact only with their immediate neighbors or the individuals with whom the players are connected. In the BA network,  such spatial constraints can lead to the evolution and maintenance of cooperation.

Well-mixed populations is another subject of study in evolutionary game theory. Here, an individual is free to interact with every other individual in the population. This situation is often modeled using replicator dynamics. A complete graph can be used to study such scenarios, with each individual being connected to every other, all the time.  However, such structures may not be suitable for the maintenance of cooperation. Also, there are some fundamental biological problems with such models. At a certain instant of time, an individual can only interact with a limited number of neighbors depending on its biological capability. As an example we could consider the extremely common scenario of a ``mixed bacterial culture" in liquid medium. The public goods produced by a bacterium at any particular instant of time would diffuse. This diffusion will be decided by the concentration profile in the medium and other factors.   Essentially, the immediate neighbors of the bacterium at that  instant are benefited but not neccessarily the entire population. Therefore, modeling well-mixed scenarios using a complete graph cannot really be the choice of preference. 

In between these two extremes, intermediate states arising from mobility, dispersal or random fluctuations could well exist \cite{chiong2012random, chen2012risk, Vishu2017}.  These fluctuations could arise due to various reasons ranging from fluctuations in the graph connectivity to diverse ecological parameters like immigration, temperature or availability of food. Mobility could have both positive or negative effect on the maintenance of cooperation. When the movement of an  individual is driven by evolutionary success, it could have a positive impact on the maintenance of cooperation \cite{Chiong2012, Cardinot2019}. Mobility has also been thought to promote cooperation in dynamic groups via emergent self-assortment dynamics \cite{Vishu2017}. However, random dispersal and fluctuations could also prove detrimental to cooperation \cite{Antonioni2011}. 

In evolutionary games, the game parameters, namely the payoffs, are generally invariant.  However, as well-known,  when such games are played on graphs, the underlying network structure can significantly influence the outcome \cite{santos2005scale}. Remarkably, in evolutionary games on networks, the dependence of payoff on the underlying  topology of the network has not really been studied. Therefore, the wider ramifications of spatial restrictions on game payoffs definitely need to be unravelled.  Herein we introduce an appropriately altered form of payoff, which incorporates the underlying topology  of the network, and, thence demonstrate their importance in games played on graphs. 

It is simultaneously important to emphasise here that the notion of a non-invariant payoff matrix is well-studied  in quantum games \cite{Piotrowski2003} as well as in evolutionary game theory literature \cite{tomochi2002, chong2006, ashlock2010, PercNJP2006, masuda2007, Turner2003, tanimoto2007} as outlined below. This is essentially because the interactions between players in evolutionary games depends on many factors. The fraction of defectors in society affects the payoff values, which proves to be necessary for describing evolution \cite{tomochi2002}. In a different co-evolutionary framework, self-adaptivity of payoff matrix has also been implemented for describing the real-world nature of the players \cite{chong2006, ashlock2010}. In addition, the interaction frequencies encountered by a player could affect the payoff matrix \cite{PercNJP2006, masuda2007}. 
Noise leading to deviations from traditional payoff values could even lead to a change in the very nature of the game itself \cite{tanimoto2007}. Mutations responsible for the inclusion of an alternate strategy could be responsible for mutualism, thereby leading to a shift of the game dynamics towards stable cooperation \cite{Worden2007}.  Thus, such  alterations in the nature of a game is well-known from both  experimental \cite{Turner2003} and theoretical studies \cite{tanimoto2007, Worden2007, Gibbons1991}.

Our approach is inspired by the phenomenon of quorum sensing \cite{bruger2016} as detailed in Sec.~\ref{sec:quorum}. In existing literature, the impact of spatial restrictions on the outcome of the game has been studied mostly via the inherent limitations imposed by the dint of these interactions rather than through the game parameters {\em per se}.  We observe that when measured in terms of topology dependent pay-offs, the game outcome demonstrates that  cooperation is supported in the presence of random dispersal. Indeed, changes can be witnessed in the very nature of the game itself.  We also show that if the initial cooperation is very high, then cooperation is most likely to remain the dominant strategy in the population. Lastly, the notion of topology dependent payoffs is important for all games on networks and not just for the case of prisoner's dilemma studied herein.

\begin{figure*}[t]
\resizebox{1\textwidth}{!}{%
	\includegraphics{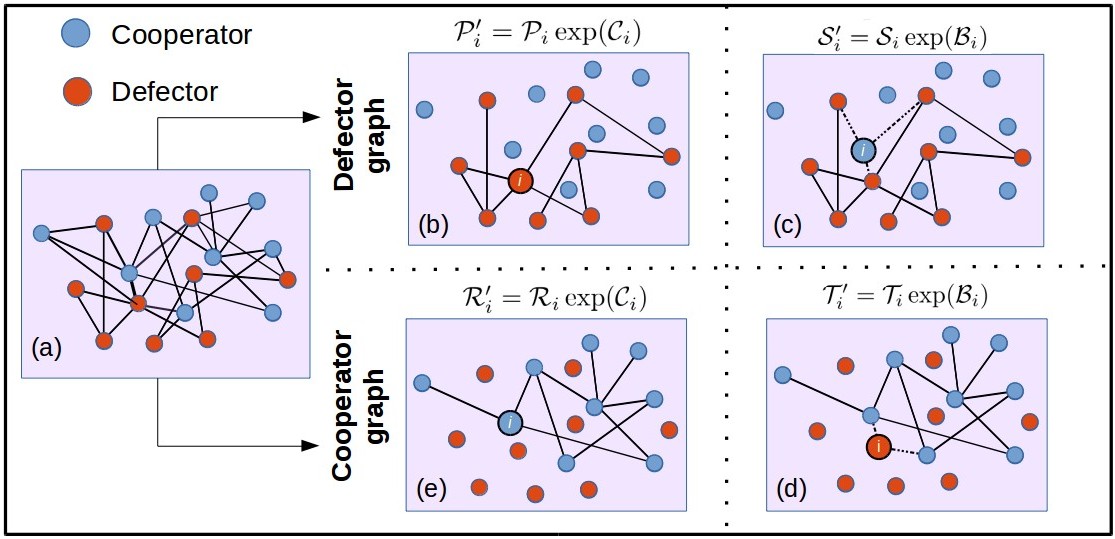}
}
\caption{Eqn.\ref{eq:a_p} quantifies the determination of topology dependent payoffs, ${ \Pi_i^{\prime}}$, for an individual, $i$. The original graph is represented in (a). The  ``defector graph" is constituted only of defectors and the links among defectors. It determines altered punishment, ${\cal P}_i^{\prime}$ and altered sucker's payoff, ${\cal S}^{\prime}$,  as represented in (b) and (c) respectively. Similarly, the ``cooperator graph" possesses only cooperators and the links among cooperators. It determines the altered temptation, ${\cal T}_i^{\prime}$, and altered reward, ${\cal R}_i^{\prime}$, as represented in (d) and (e) respectively. The solid lines signify intra-species interactions in (b), (c), (d), and (e).  The dotted lines represent the inter-species interactions of $i$ in (c) and (d)}
	\label{fig:Cgraph} 
\end{figure*}

\section{Model}
\subsection{Algorithm}
We  follow an extremely well-studied game algorithm \cite{santos2005scale} for simulating evolutionary PD game in structured populations. Here, each of the $N$ individuals represent a node on a BA network. 
According to this algorithm, an individual can interact with all of its neighbors depending on the population structure. At every round of game, the individual will accumulate payoff according to the game rules. As widely studied in earlier literature,  following values of the PD game parameters are adopted, $1<{\cal T}\le2$, ${\cal R}=1$ and ${\cal P}={\cal S}=0$. 

After the determination of payoff, the strategy of each individual is updated synchronously. We have considered imitation as the mechanism for strategy upgradation in our simulations \cite{ohtsuki2006replicator}, as detailed below. For the strategy upgradation of an individual, $i$; one of its neighbors, $j$, is randomly selected.  Their respective payoffs $\Pi_i$ and $\Pi_j$ are compared. $i$ upgrades its strategy whenever $\Pi_{j} > \Pi_{i}$ with a probability of $(\Pi_{j}-\Pi_{i}) / (Z{k^*})$. Here $Z = T-S$  and $k^{*} = max (k_i , k_j )$. Here $k_i$ and $k_j$ is the degree of $i$ and $j$ respectively. After a transient time of $10^{4}$ generations, the final number of cooperators, $f_C$, is counted over $10^{3}$ generations. 
The overall simulations are repeated for each value of the relevant game parameter for an ensemble of $E_N$ networks. The networks have $N$ nodes, where each node possesses an average degree $\langle k \rangle$. Data has been collected in the range ${\cal T} = [ 1.01, 1.91]$ at equal intervals.  $f_{C_i}$ is the initial fraction of nodes, which are randomly assigned to be cooperators at the start of the simulations. Obviously $(1 - f_{C_i})$ is the initial fraction of nodes, which are defectors.

\subsection{Static population with random dispersal}
The literature on evolutionary games predominantly considers populations, where all players possess fixed spatial positions. The number of interactions that every player participates in, depends on the structure of the static graph. For graphs with a low average degree, the presence of spatial restrictions  does not allow a player to interact directly with every other. 
Some studies incorporate a higher average degree of graphs to model well-mixed scenarios \cite{santos2005scale}. The extreme case is obviously that of a complete graph, where each individual is connected with every other. However,   as aforementioned,  this has obvious biological limitations.

In our model, random dispersal, $\mu$, has been incorporated as a probability that player, $i$, interchanges its strategy with another randomly chosen player \cite{Antonioni2011}.  We have studied various levels of $\mu$ in our simulations. Higher dispersal would mitigate the maintenance of cooperation  to a great extent. Herein the players possess pure strategies, namely cooperation and defection. Therefore, the interchange of strategy between two randomly chosen players  also signifies that these players are interchanging their position. However, {\em the degree sequence of the network is maintained intact}, during the entire process. This is essential because dispersal should not be achieved at the cost of changing the network structure itself. 

\begin{figure}[htbp]
\resizebox{0.485\textwidth}{!}{%
\includegraphics{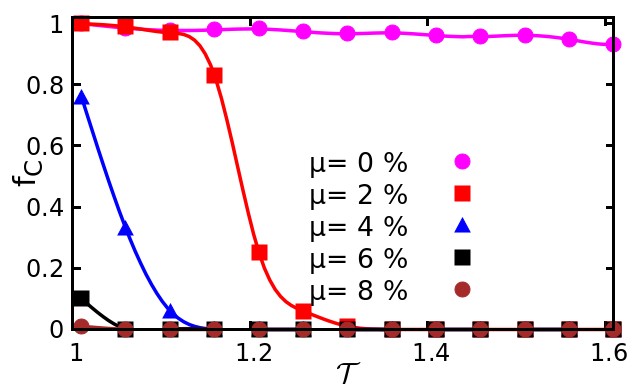}%
}
\caption{\label{fig:dynamic} Fraction of cooperators, ${ f}_{{\cal C}}$, versus temptation, ${\cal T}$, in the absence and presence of random dispersal, $\mu = 2\%, 4\%, 6\%$ and $8\%$. Here, the initial fraction of cooperators, $f_{C_i}=0.5$.  Results are for $N=1024$, $ \langle k\rangle=4$ and $E_N = 200$ ensembles. Evidently, increase in $\mu$ leads to loss of cooperation. The standard error is smaller than the size of the data points.}
\end{figure}

\subsection{Quorum sensing}
\label{sec:quorum}
The study of game theory \cite{Lotem2003, Diggle2007,Traulsen2007} including prisoner's dilemma \cite{iliopoulos2010}, in quorum sensing and allied phenomena is well-known in literature. PD is quite useful for studying the control exercised by individual players over public goods \cite{schuster2017}. In many bacterial populations, quorum sensing is thought to play an important role in the maintenance of cooperation \cite{Czaran2009, Requejo2011}. 

{\em We should emphasise that herein we do not purport to model quorum sensing in any manner. Instead, we seek to implement lessons from quorum sensing in our model.} The central notion of quorum sensing is that the benefits provided by a group of individuals are more readily available to the members of an alliance rather than to the rest of the populace. 

In our model -- we consider the entire  population of cooperators, in form of the cooperator graph -- in order to determine the payoff of a single cooperator. Similarly, the entire  population of defectors -- in form of the defector graph -- is considered in order to determine the payoff of a single cooperator. The lesson from quorum sensing is that it is not merely the local neighborhood which always decides the outcome. This is schematically shown in Fig.~\ref{fig:Cgraph}. The detailed method of such calculation of these payoffs is discussed in Sec.~\ref{sec:topology-payoff}

\subsection{Topology dependent payoffs}
\label{sec:topology-payoff}

In general, payoffs are considered to be invariable. But  interactions between the players might depend on many factors \cite{Requejo2012}.  In most of existing literature, the accumulated payoff of an individual depends on the payoffs of its immediate neighbors. However, in any network, an individual will interact not only with its immediate neighbors but also with distant individuals  connected indirectly to it through various shortest paths.

Here we are interested in the evolution and maintenance of cooperation in the presence of random dispersal. Thus interaction between any two players is quite likely to be affected by their own connections with the remaining players \cite{szolnoki2017environmental}. Therefore, should the reward acquired by a typical cooperator, $C$, be  determined merely by those of its nearest cooperator neighbors? We demonstrate the consequences of taking  into consideration other cooperators in the population, who are not the immediate neighbors of $C$. Indeed, we show that they can significantly  influence the reward acquired by $C$.  As aforementioned, the idea of including all cooperators in a network -- instead of merely the nearest cooperator neighbors -- is influenced by quorum sensing.

We now introduce the concept of the ``cooperator graph", ``defector graph" and topology dependent payoffs as depicted in Fig. \ref{fig:Cgraph}. The cooperator graph, $\cal G_{C}$, is a {\em virtual construct}, achieved simply by pruning every defector, $D$, and all connections of $D$ from the original graph, $\cal G (V,E)$. Here, $\cal {V}$ and $\cal E$ represent the set of nodes and edges respectively in $\cal G$. All edges between $D-D$ and $C-D$ have been omitted from $\cal G (V,E)$, while all $C-C$ edges are exclusively retained. Clearly, $\cal G_{C}$ is then a graph of  cooperators only. Similarly, the ``defector graph", $\cal G_{D}$, can be obtained by omitting the edges of the type $C-C$ and $C-D$. 

Throughout this paper, ${\cal C}_{i}$ and  ${\cal B}_{i}$ respectively denote the closeness centrality and betweenness centrality of node, $i$, in the network.  The importance of network centrality measures, particularly ${\cal C}_i$, and ${\cal B}_i$, in diverse contexts like optogenetics, image processing, non-invasive diagnostics and game theory itself is well-known  \cite{roy2009, banerjee2015, kaur2015, antonioni2013coordination, banerjeeSR2015, deb2020residue}.  The importance of the shortest paths in evolutionary algorithms for collecting public goods has also been studied previously \cite{cho2009}. ${\cal C}_i$, and ${\cal B}_i$, have also found applications in ecological networks and social networks \cite{Jordan2007, Tylianakis2018}.

Let us focus on intra-species interactions i.e., in ``cooperator graphs" or ``defector graphs". 
As mentioned above, a given pair of  individuals who are not immediate neighbours can be connected through the shortest paths joining them. The lower the distance between two individuals, the  higher would be the communication expected between them and vice-versa. Hence, their influence on each other's payoff should also vary accordingly. This situation can be aptly addressed with the help of closeness, ${\cal C}_{i}$. The closeness of a node indicates how ``close" the node is to the remaining nodes of the network. A cooperator, who is closer to other cooperators in the cooperator graph possesses the luxury to enjoy higher public goods. ${\cal C}_i$, of the node, $i$, is the normalized reciprocal of the total sum of shortest path lengths between $i$ and all other nodes in a network of $N$ nodes. Thus,
\begin{equation}
{\cal C}_i = \frac{N-1}{\sum_{j=1}^{N-1} d(j,i)},  
\label{eq:closeness}
\end{equation}
where, $d(j,i)$ is the shortest path between node $i$ and every other node $j$ in the network. 

In inter-species interactions, the altered temptation of a given defector depends on its interactions with all cooperators of that network. These interactions can be easily understood by observing how that defector is connected with the cooperator graph. We can similarly calculate the sucker's payoff of a given cooperator, depending on its connections with other defectors in the graph. For this, we consider the links of this cooperator with the defector graph. Therefore, an individual of one species can enjoy public goods due to other species, affecting its payoff. Hence, the importance of betweenness arises in the case of inter-species interactions. It signifies how an individual interacts with others belonging to a different species. 
%Betweenness, 
${\cal B}_i$, of a node, $i$, is the fraction of shortest paths passing through $i$, compared to all pairs of shortest paths in the network. Thus, 
\begin{equation}
{\cal B}_i = \sum_{i\neq j\neq k} \frac{\sigma_{jk} (i)}{\sigma_{jk}}, 
\label{eq:betweenness}
\end{equation}
where, $\sigma_{jk}$ denotes the total number of shortest paths between nodes $j$ and $k$, with $i\ne j\ne k$. $\sigma_{jk} (i)$ represents only those shortest paths of $\sigma_{jk}$, which pass through $i$.

We now propose a general form of the topology dependent payoff for an individual, $i$, as
\begin{equation}
{ \Pi_i^{\prime}} ={\Pi_i} \exp (a {\cal C}_{i}+b {\cal B}_{i})
\label{eq:a_p}
\end{equation}
$\Pi_i$ is the conventional form of the payoff \cite{santos2005scale, Szolnoki2009}.  $a$ and $b$ are the parameters representing intra-species and inter-species interactions, respectively. For intra-species interactions ($C-C$ and $D-D$); $a=1$ and $b=0$. Again, in case of inter-species interactions ($C - D$ and $D - C$); $a=0$ and $b=1$. The choice of the altered form of payoff does not cause any loss of generality, associated with the conventional payoff matrix \cite{Szabo2007}. The altered form of payoff incorporates the topology of the underlying structure. 
 An individual will accumulate ${\Pi}_{i}^{\prime}$ instead of $\Pi_i$ while interacting with all of its neighbors. Depending on the strategy of the interacting individuals, ${\Pi}_{i}^{\prime}$ could be $\cal T'$, $\cal R'$, $\cal P'$ or $\cal S'$, as illustrated in Fig. \ref{fig:Cgraph}.  Indeed, as the nature of the topology changes from random to regular, clearly $\Pi^{\prime} \to \Pi$.
\begin{figure}[htbp]%
%\centering%
\hspace{-0.01in}%
\resizebox{0.485\textwidth}{!}{%
	\includegraphics{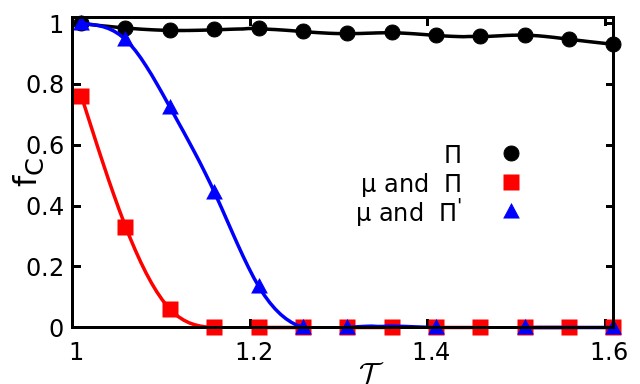}
}
\caption{\label{fig:dynamic2} Fraction of cooperators, ${ f}_{{\cal C}}$, versus temptation, ${\cal T}$, in the presence and absence of $\mu$. The initial fraction of cooperators, $f_{C_i}=0.5$, and, $ \langle k\rangle=4, \mu = 4\%$, $E_N = 200$, and $N=1024$. In the presence of random dispersal, cooperation is witnessed to some extent, when measured in terms of the topology dependent payoff, $\Pi^\prime$, rather than the conventional payoff, $\Pi$. Standard error is smaller than the size of data points.}%
\end{figure}%

\begin{figure}[htbp]%
%\centering%
\hspace{-0.01in}%
\resizebox{0.485\textwidth}{!}{%
	\includegraphics{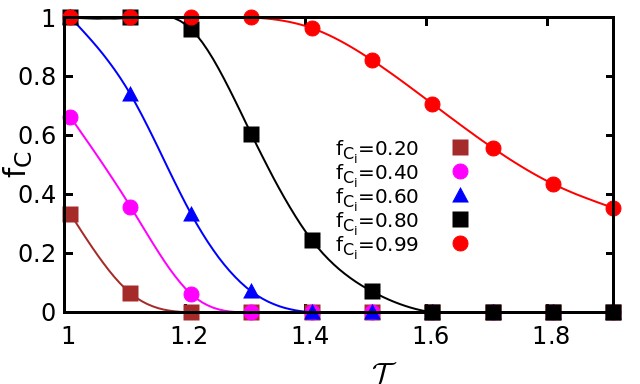}
}
\caption{\label{fig:dynamic5} The fraction of cooperators, $f_C$, at the end of the simulations with respect to  temptation, ${\cal T}$, at various initial fraction of cooperators, ${f}_{{C}_i}$,  in the presence of random dispersal and $\Pi$. Results are for $N=1024$, $ \langle k\rangle=4$, $\mu = 4\%$ and $E_N = 200$. Standard error is smaller than the size of data points.}%
\end{figure}%

\begin{figure}[htbp]%
%\centering%
\hspace{-0.01in}%
\resizebox{0.485\textwidth}{!}{%
	\includegraphics{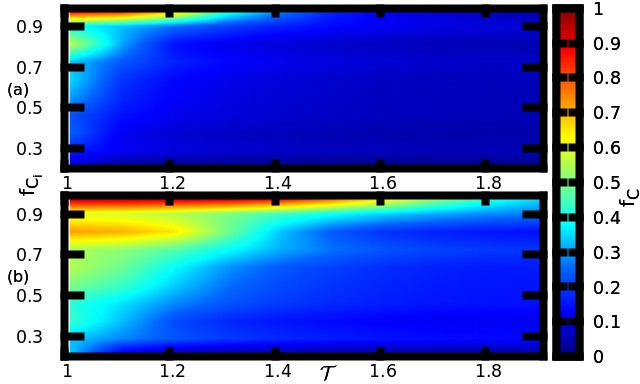}
}
\caption{The variation of $f_{C_i}$ versus ${\cal T}$  at various values of $f_C$, as quantified by: (a) conventional payoff, $\Pi$, and, (b) topology dependent payoff, $\Pi^{\prime}$. Red denotes maintenance of cooperation and blue its absence. Significant enhancement of cooperation in presence of random dispersal can be observed, when measured in terms of $\Pi^{\prime}$ rather than $\Pi$. Results are for $N=1024$, $ \langle k\rangle=4$, $\mu = 4\%$ and $E_N = 200$.}%
\label{fig:dynamic7}%
\end{figure}%

We consider the cooperator and defector graph for the calculation of altered reward, ${\cal R}_i^{\prime}$, and altered punishment, ${\cal P}_i^{\prime}$, respectively. For ${\cal R}_i^{\prime}$ and ${\cal P}_i^{\prime}$, we consider only the intra-species interactions. Thus, the form of the topology dependent payoffs are ${\cal R}^{\prime}_i={\cal R}_{i}\exp({\cal C}_{i})$ and ${\cal P}^{\prime}_i={\cal P}_{i}\exp({\cal C}_{i})$ respectively. On the other hand, for the altered payoffs, ${\cal T}_i^{\prime}$ and ${\cal S}_i^{\prime}$, the inter-species interactions are considered.The  form of altered temptation and altered punishment is ${\cal T}^{\prime}_i = {\cal T}_i \exp({\cal B}_{i})$ and ${\cal S}^{\prime}_i = {\cal S}_i \exp({\cal B}_i)$ respectively. The determination of ${\cal P}^{\prime}_i , {\cal S}^{\prime}_i, {\cal T}^{\prime}_i$ and ${\cal R}^{\prime}_i$ for a node, $i$, is depicted in Fig. \ref{fig:Cgraph}(b), Fig. \ref{fig:Cgraph}(c), Fig. \ref{fig:Cgraph}(d) and Fig. \ref{fig:Cgraph}(e) respectively. 

In summary, the value of topology dependent payoffs depend on the spatial position of the player, and more explicitly on the player's centrality in the network. {\em However, we can hardly overemphasise that more than the precise form of the payoffs,  herein we want to stress on the importance of incorporating the effect of the underlying topology on the game payoffs.}

\begin{figure}[htbp]
\hspace{-0.01in}%
\resizebox{0.485\textwidth}{!}{%
	\includegraphics{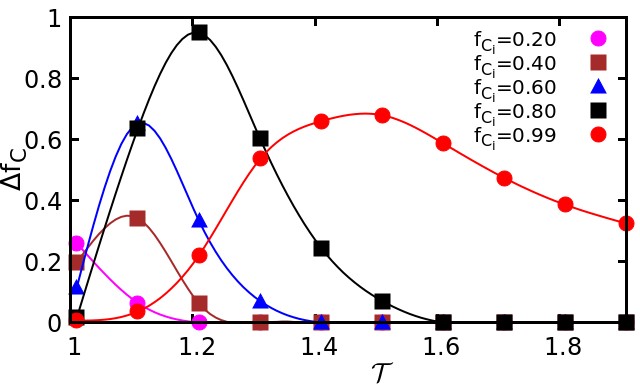}
}
\caption{Plot of $\Delta {f_C}=(f_{C_{{\Pi}^\prime}} - f_{C_{\Pi}})$ versus $\cal T$. Here, $f_{C_{\Pi}}$ and $f_{C_{{\Pi}^\prime}}$  denote the cooperator fraction in the absence and presence of altered payoff, ${{\Pi}^\prime}$, respectively. Results are for $N=1024, \langle k\rangle=4, \mu =4\%$ and $E_N = 200$.}
\label{fig:dynamic6} 
\end{figure}

\section{Topology dependent payoffs: Alteration of game dynamics}
In the absence of random dispersal, cooperation obviously dominates the population due to the presence of hubs \cite{santos2005scale}. Fig. \ref{fig:dynamic} presents the behaviour of $f_{C{_i}}$ versus $\cal T$ at various values of random dispersal, namely, $\mu = 0, 2\%, 4\%, 6\%$ and $8\%$. With  increasing $\mu$, cooperators are unable to accumulate higher payoffs. Increasing $\mu$ leads to a random change in the strategies of players and the cooperator fraction goes down with increase in $\cal T$. For obvious reasons, we have chosen an intermediate value of $\mu = 4\%$ for our remaining simulations, rather than choosing a higher or lower value of $\mu$. 

Fig. \ref{fig:dynamic2} demonstrates the behaviour of cooperation, when measurement is done using topology dependent payoff, $\Pi^\prime$, as compared to the conventional payoff, $\Pi$.  We observe that when measured by $\Pi^\prime$, cooperation is maintained to some extent, irrespective of $\mu$. 

We have also investigated the effect of initial cooperator fraction at the start of the simulations, $f_{C_{i}}$, on  maintenance of cooperation. In Fig. \ref{fig:dynamic5}, we observe the  behavior of $f_C$ versus ${\cal T}$ at different values of $f_{C_{i}}$. Clearly, only rather higher values of $f_{C_i}$ can lead to the maintenance of cooperation at higher ${\cal T}$. 

In Fig. \ref{fig:dynamic7}, we compare $f_{C}$ at different values of $f_{C_{i}}$ and $\cal{T}$, when measured by both conventional and topology dependent payoff at different values of $f_{C_i}$ at $\mu = 4\%$. The importance of measuring by altered payoff becomes immediately clear. The maintenance of  cooperation upon consideration of topology dependent payoffs is much higher in comparison to the scenarios, where they are not considered.  To  understand this further, we  create a ``differential plot". $f_{C_{\Pi}}$ and $f_{C_{{\Pi}^\prime}}$  denote the cooperator fraction in the absence and presence of ${\Pi}^\prime$ respectively. $\Delta {f_C}=(f_{C_{{\Pi}^\prime}} - f_{C_{\Pi}})$ has been plotted with respect to $\cal T$ in Fig. \ref{fig:dynamic6} at various $f_{C_i}$. A decreasing  slope implies loss of cooperation, when measured by topology dependent payoffs. On the other hand, a positive slope seems to signify the tendency of cooperators to maintain their population. We can observe the peaks shifting towards higher $\cal T$, with an increase in $f_{C_i}$. This indicates the prevalence of  cooperation at higher $f_{C_i}$ even in the presence of $\mu$, when measured by ${\Pi}^\prime$.

\begin{figure}[htbp]%
%\centering%
\hspace{-0.01in}%
\resizebox{0.48\textwidth}{!}{%
	\includegraphics{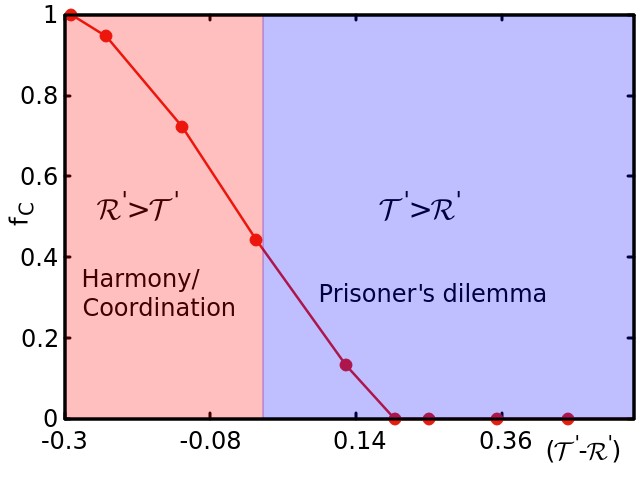}
}
\caption{\label{fig:dynamic3}Fraction of cooperators, $f_C$, versus $({\cal T}^{\prime}- {\cal R}^{\prime})$. Results are for $N=1024$, $ \langle k\rangle=4$, $\mu=4\%$, $E_N = 200$  and $f_{C_i} = 0.5$. In the shaded region, ${\cal R}^{\prime} > {\cal T}^{\prime}$, which indicates that players are not participating in PD anymore. Therefore, the players engage in either harmony or coordination game, depending  on the value of $\cal S^\prime$ and $\cal P^\prime$ \cite{Szabo2007, Sinha2019}.}%
\end{figure}%

%\vspace{-0.1in}
As aforementioned, according to the basic rules of the Prisoner's Dilemma game, $\cal T > R$ and $1<{\cal T}\le 2$.  In the present scenario, the average value of altered temptation $1<{\cal T}^{\prime} \le 2$. Remarkably, however in Fig. \ref{fig:dynamic3}, we observe ${\cal R}^{\prime} > {\cal T}^{\prime}$ in a certain region. This region would obviously seem to violate the basic condition of PD. However, experiments on bacteriophages have shown that ``escaping" PD  is  well known in literature \cite{Turner2003}. Herein, drawing an analogy with Ref.  \cite{Turner2003}, altered reward helps the cooperators in escaping the  PD game and the associated strategy of commons. The cooperators  engage themselves in either Harmony or Coordination games \cite{Sinha2019}, when measured in terms of the altered reward. Similar instances of changes in the nature of the game is well-known in  literature \cite{tanimoto2007, Piotrowski2003}.

\section{Conclusion}
Does the underlying topology of the graphs on which games are played have no role whatsoever for measuring the dynamics and outcome of the games? Non-invariant payoffs have been well-studied in game theory \cite{tomochi2002, chong2006, ashlock2010, PercNJP2006, masuda2007, Turner2003, tanimoto2007}. Therefore, inspired by quorum sensing, herein we propose an appropriately altered form of payoff which is dependent on the underlying topology of the network. Even if the topology dependent payoffs possess a weak dependence on the underlying topology, significant change in the outcome and indeed in the nature of the game itself is witnessed. Furthermore, cooperation can be maintained even in the presence of random dispersal. Such changes in the nature of the game are known from both experimental \cite{Turner2003} and theoretical studies \cite{tanimoto2007, Worden2007, Gibbons1991}. Future studies can unravel a more precise and sensitive dependence of the payoff on network topology. As well known, mean-field-type phase transitions on random regular graphs and regular small world networks have been observed in games on structured populations. It would be interesting to examine such behaviour with topology dependent payoffs. 
\section{Authors contributions}
All the authors were involved in the preparation of the manuscript. All the authors have read and approved the final manuscript.
%
% BibTeX users please use
 \bibliography{quorum}
 \bibliographystyle{epj}
\end{document}